# Tuning Coherent-Phonon Heat Transport in LaCoO$_3$/SrTiO$_3$ Superlattices.

D. Bugallo,[1] E. Langenberg,[2] E. Carbó-Argibay,[3] Noa Varela Dominguez,[1] A. O. Fumega,[4,5] V. Pardo,[4] Irene Lucas,[6] Luis Morellón,[6] F. Rivadulla.[1,*]

[1]*Centro Singular de Investigación en Química Biolóxica e Materiais Moleculares (CIQUS), Departamento de Química-Física, Universidade de Santiago de Compostela, 15782 Santiago de Compostela, Spain.*

[2]*Department of Condensed Matter Physics, Institute of Nanoscience and Nanotechnology (IN2UB), University of Barcelona, Spain.*

[3]*International Iberian Nanotechnology Laboratory (INL), Av. Mestre José Veiga s/n, 4715-330 Braga, Braga, Portugal.*

[4]*Departamento de Física Aplicada, Universidade de Santiago de Compostela, 15782 Santiago de Compostela, Spain.*

[5]*Department of Applied Physics, Aalto University, FI-00076 Aalto, Finland.*

[6]*Instituto de Nanociencia y Materiales de Aragón (INMA), Universidad de Zaragoza and Consejo Superior de Investigaciones Científicas, 50009 Zaragoza, Spain.*

**Abstract.** Accessing the regime of coherent phonon propagation in nanostructures opens enormous possibilities to control the thermal conductivity in energy harvesting devices, phononic circuits, etc. In this paper we show that coherent phonons contribute substantially to the thermal conductivity of LaCoO$_3$/SrTiO$_3$ oxide superlattices, up to room temperature. We show that their contribution can be tuned through small variations of the superlattice periodicity, without changing the total superlattice thickness. Using this strategy, we tuned the thermal conductivity by 20% at room temperature. We also discuss the role of interface mixing and epitaxial relaxation as an extrinsic, material dependent key parameter for understanding the thermal conductivity of oxide superlattices.

TOC Graphic

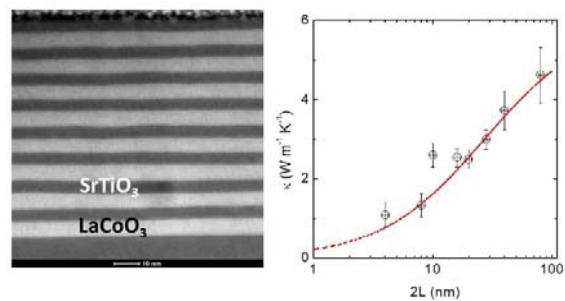

There are three important length scales whose relative size determines the lattice thermal conductivity, κ, in nanostructures: the phonon mean free path, ℓ, their wavelength, λ, and a characteristic physical length of the system, L (the period length in a multilayer, for instance).[1] For periodically arranged interfaces, as in a super-lattice (SL), phonons of sufficiently long-wavelength (λ > L) may undergo wave-interference effects, given their ℓ is long enough to propagate over several interfaces (*i.e.* ℓ > L). In this regime, phonons behave as coherent waves, with a ω(k) dispersion characteristic of the SL, with their group velocity and density of states for each polarization, as well as energy gaps that forbid the propagation of certain phonon frequencies, decreasing κ of the SL as L increases.[2] Coherent propagation of thermal phonons was demonstrated by Luckyanova *et al*.[3] in GaAs/AlAs SLs, taking advantage of the long λ and ℓ in these semiconductors. Control of κ through wave-interference effects has been also achieved in Si nanostructures with periodically arranged patterns, spaced ≈ℓ.[4–6] Note that the high sensitivity of wave-interference effects to the periodicity of the SL, introduces another tunable parameter to control the thermal conductivity of a nanostructure, at a length scale that should not affect much the electrical conductivity, raising the interest for thermoelectric applications.[7]

On the other hand, a progressive increase of L will put more phonons at ℓ < L, increasing the contribution from incoherent phonons to κ.[8–12] Thus, the crossover from a regime in which heat transport is governed by coherent phonons, to another one in which incoherent phonons dominate, should, in principle, be signaled by a minimum in the thermal conductivity of the SL at a given L.[2,13]

Most of the experimental work to corroborate this crossover has been carried out in semiconductor SLs, due to their large mean free path, the possibility of growing clean interfaces (defects of the order of λ produce diffuse reflections and loss of phonon-phase coherence, resulting in particle-like, incoherent, propagation), and their interest in thermoelectric applications.

For instance, a minimum at κ (L) was reported by Chakraborty *et al*.[14] in Si-Ge SLs at L≈7 nm; Venkatasubramanian[15] also observed the crossover in $Bi_2Te_3/Sb_2Te_3$ superlattices, at L≈ 5 nm. However, other authors did not find evidence of the minimum at κ(L) in $Bi_2Te_3/Sb_2Te_3$ or GaAs/AlAs SLs, suggesting a critical effect of interface roughness.[16–19] Comparing the results from GaAs/AlAs SLs of different thickness and periods, Cheaito *et al*.[18] confirmed the contribution of both incoherent and coherent phonons, even in the absence of the minimum κ (L); similar conclusions were reached by Luckyanova *et al.*,[20] and Alaie *et al.*,[6] the latter in Si membranes.

Regarding oxide multilayers, a shallow minimum at L=2-3 nm was reported by Ravichandran *et al*.[21] in SLs of $SrTiO_3/CaTiO_3$ and $SrTiO_3/BaTiO_3$. On the other hand, Katsufuyi *et al*. observed a linear increase of κ(L) in $SrTiO_3/SrVO_3$ SLs, from L≈4-100 nm,[22] without any sign of κ(L) minimum or flattening at low L. Instead, these authors reported a constant interfacial resistance ≈2 × $10^{-9}$ $Km^2/W$. A monotonic decrease was also observed in the thermal conductivity of $(SrTiO_3)_nSrO$ Ruddlesden–Popper superlattices, as the interface density increases.[23]

Transition-metal 3*d* oxides are quite ionic, and sharp interfaces may introduce polar discontinuities in some cases, whose energy penalty can be resolved through ionic intermixing.[24] This, together with their shorter ℓ and λ than semiconductors, should make them more sensitive to interfacial defects.[25,26] Thus, the observation of a minimum κ(L) in oxide SLs should be more difficult than in semiconductors, and places the question of how relevant wave-interference effects are in oxide SLs, and how much their thermal conductivity may be tuned acting over coherent-phonons.

Here we report a systematic study of $LaCoO_3/SrTiO_3$ (LCO/STO) SLs, varying the total thickness *t*, and the lattice period, L, as well as the periodicity of the structures. We show that both coherent and incoherent phonons contribute to κ of the SL, at all periodicities. We also demonstrate that the contribution of long-wavelength coherent phonons can be reduced by the effect of small variations of the periodicity, even at large L and at room temperature, which could be useful in thermoelectrics and in thermal management devices.

A series of $LaCoO_3/SrTiO_3$ (LCO/STO) SLs with a total thickness of *t*=80 nm and different periods were synthesized by PLD (see Table S1, supporting information). We denoted our SLs by (L×n), where L is the thickness of each individual layer forming the SL (so that the period of the superlattice is 2L) and n is the repetition of each layer. The materials for the SL of this study, STO, $a_{STO}$=3.905 Å, and LCO, $a_{LCO}$=3.80 Å, were selected, under the premise of having different enough masses and lattice constants, but still show a good epitaxial growth on top of each other, to have good crystallinity and well-defined interfaces. The cumulative thermal conductivity calculated ab-initio, Figure S1 in the supporting information, shows that phonons have similar mean free paths in STO and LCO. Also, from these data, a substantial effect is expected for periods 2L≈10-20 nm, between 100-250 K. Thus, LaCoO$_3$ and SrTiO$_3$ seem a good compromise for studying the contribution of coherent/incoherent phonon transport in epitaxial oxide SLs.

The structural and microstructural analysis of the samples is summarized in Figure 1. The Θ-2Θ X-ray diffraction patterns around the (002) peak of the STO substrate show an increasing number of periodically spaced SL peaks as the number of periods increases, indicating the preservation of the long-range order. The satellite peaks along the Q$_z$ axis in the reciprocal space maps (RSM, Figure 1 b) further confirm the periodic structure along the out-of-plane direction of the SL and show that the SLs are coherently strained with the substrate—see also the discussion of the scanning transmission electron microscopy (STEM) data below. X-ray reflectivities show the characteristic SL peaks, as well as a smooth angular decay, suggesting smooth interfaces. The thickness of the SLs obtained from these analyses are in good agreement with each other and correlate very well with the nominal thickness (see table S1 in the supporting information, and Figure S2, for further details of the fittings and results of the structural analysis).

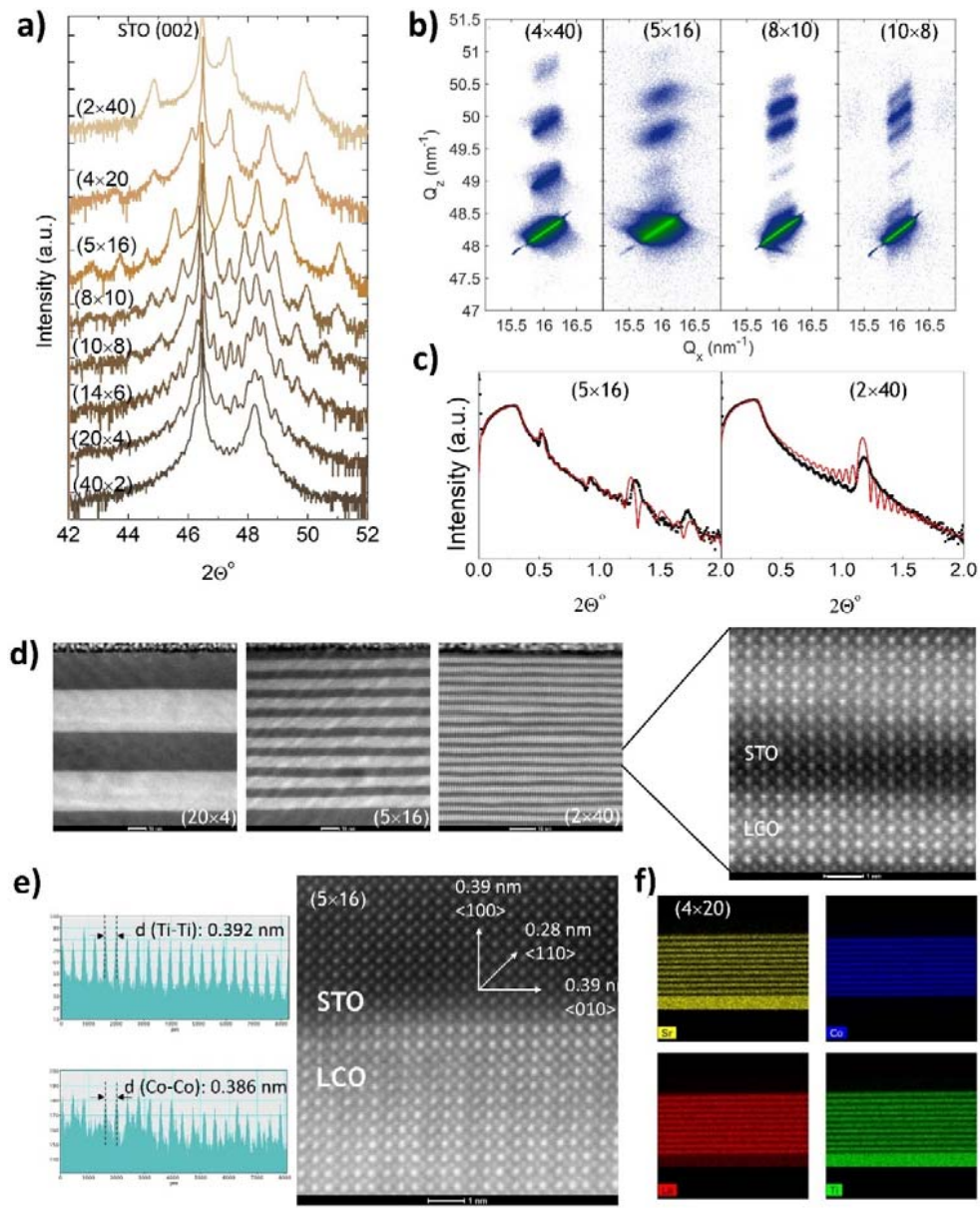

**Figure 1.** a) Θ-2Θ X-Ray diffraction pattern of the SLs around the (002) peak of the (001)-oriented STO substrate. The total thickness of each SL is $t\approx 80$ nm; the periodicity is indicated in each case. b) RSM around the (103) reflection of the STO substrate. The periodicity is indicated in each panel. c) X-Ray reflectivity of two samples showing the SL peaks and smaller oscillations related to the total thickness. d) Cross section high-angle annular dark-field (HAADF)-STEM images of several SLs, with the period indicated on each panel. The scale (white bar) is 10 nm in every image, except in the zoomed area of the (2×40) SL, right, which is 1 nm. e) Image intensity profiles (displaying Sr-Sr and Co-Co spacings) parallel to the plane of the sample, and HAADF-STEM image from a cross-section lamella of a (5×16) SL, showing the crystalline structure. The metal-metal distance obtained from the image intensity profiles analysis is ≈3.92 Å for Ti-Ti (STO) and ≈3.86 Å for Co-Co (LCO), denoting a slight relaxation in STO. f) EDX map analysis of a lamella from a (4×20) SL, showing the regularity of the layer thickness. The total thickness of the SL is ≈79 nm, giving an average of 3.95 nm per layer, very close to the intended 4 nm per layer.

The microstructure of the internal interfaces was studied by high resolution STEM on several cross-section lamellae of different SLs. The results (Figure 1 d)-f), show that LCO and STO grow epitaxially on top of each other, with very-well defined interfaces, and with a thickness very close to the nominal ones (see also Figure S2 and S3 in the supporting information). The width of the interfaces, defined as the region where the intensity of the EDX peaks (Ti and Co) decays at half its maximum value, is of the order of one-two unit cells (supporting information Figure S3). Therefore, from the X-Ray diffraction and STEM analysis, we conclude that the LCO/STO SLs present an excellent crystallinity and clean interfaces, free from a substantial number of defects that could affect our analysis of their intrinsic thermal conductivity.

The cross-plane κ(T) of the SLs was measured from 25 to 290 K by the 3ω method [27] (see supporting information for details of the measurements); the results are shown in Figure 2 for the SLs with total thickness $t$=80 nm. In all SLs, κ(T) first increases up to 150-200 K, before reaching a plateau, and then decrease slightly until room temperature in the SLs with thicker periods.

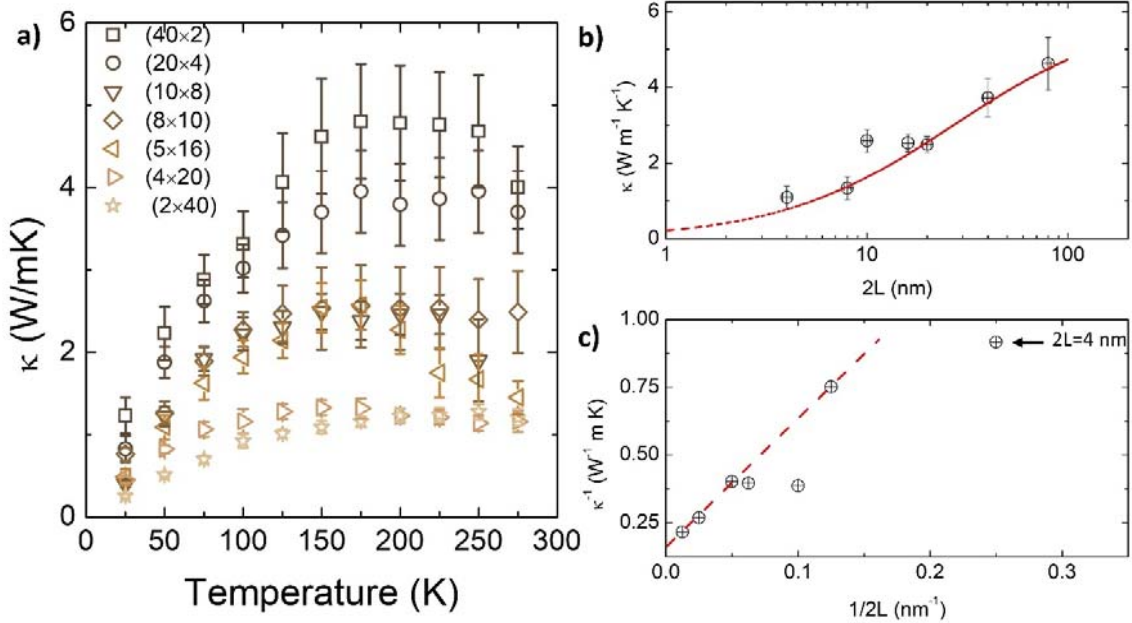

**Figure 2.** a) Temperature dependence of the thermal conductivity of the different SLs with t=80 nm (see Table S1 in the supporting information for further details of the periodicity of each sample). b) Dependence of the thermal conductivity of the SLs at 150 K with the period, 2L, and the fitting to equation (1). c) Linearized version of equation (1), and linear fitting of the experimental data. As discussed in the text, the validity of equation (1) at low values of L is compromised, when the Kaptiza and the period lengths, become comparable. For that reason, the data at 2L=4 (marked with an arrow), was excluded from the fitting.

In Figure 2 b)-c) we present the cross-plane κ at 150 K vs the SL period, 2L: the thermal conductivity of the SL decreases as 2L does, indicating that ℓ must be comparable to the SL period. Therefore, at least for a significant portion of phonons, the effect of the interface boundary resistance can be captured by a simple model incorporating the interfacial Kapitza resistance, $R_{if}$, into the Fourier's law of heat conduction across the SL:[8,9]

$$\frac{\kappa}{\kappa_0} = \frac{1}{1 + \frac{2R_{if}k_0}{2L}} \qquad (1)$$

$\kappa_0$ and $R_{if}$ represent the thermal conductivity of the bulk, free of interfaces, and the interfacial thermal resistance, respectively. This equation predicts a linear relationship between 1/κ and 1/2L; however, as shown in Figure 2b), the data for 2L below ≈20 nm, deviate progressively from this behavior. Fitting the data for 2L>10 to equation (1) gives $\kappa_0 \approx 6.25(8)$ W m$^{-1}$K$^{-1}$, close to the average of STO and LCO at this temperature (see Figure S4-S6 in the supporting information for the thermal conductivity of individual LCO and STO thin films, as well as for a short discussion of the accuracy of the thermal conductivity measurements), and $R_{if} \approx 4.7(2) \times 10^{-9}$ W$^{-1}$ m$^2$K, similar to other oxoperovskite artificial interfaces, grain boundaries, or ferroelastic domain walls.[22,28,29] Note, however, that the Kapitza length, $L_K = R_{if} k_0 \approx 24$ nm, becomes comparable to, or even larger than the period length for SLs at 2L < 20 nm; below this limit, the applicability of equation (1) is not justified. Instead, the reduction of the period thickness makes ℓ > 2L for an increasing population of phonons, so they become less sensitive to the periodicity of the internal interfaces of the SL. In this case, a wave-like treatment is probably more appropriate and κ is determined by wave interference and boundary scattering at the external interfaces of the SL. In fact, below 2L<20 nm, there is a departure from the prediction of equation (1), manifested as a plateau around 2L=10-20 nm, and a larger than expected κ for 2L=4 nm (better appreciated in the linearized plot of Figure 2c). In any case, an actual minimum in κ(2L) is not observed.

As discussed before, atomic intermixing at the interfaces of thinner period SLs could be large enough to produce diffusive scattering. Although electron microscopy showed sharp interfaces, roughness is a statistical quantity, and STEM only probes a very limited region (nanometer size) of a much wider sample (mm size). Therefore, to further asses the quality of the interfaces we measured the bulk magnetic properties of the SLs.

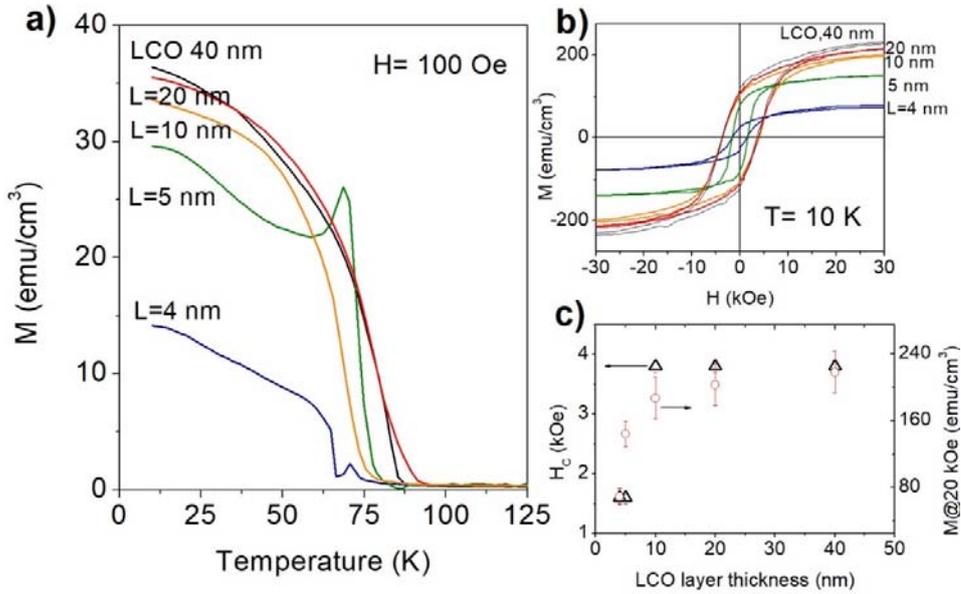

**Figure 3.** Temperature dependence of the magnetization measured at a magnetic field H=100 Oe (a), and hysteresis loops at 10 K (b) of several LCO/STO SLs; a thin film of LCO (40 nm) on STO, is also shown for comparison. c) Coercive field (triangles), $H_C$, and magnetization at 20 kOe (cricles), both obtained from the hysteresis loops in (b); there is a rapid decrease of both magnitudes in the films with L<10 nm.

Tensile strained LCO develops a magnetic order below a $T_C \approx 85$ K, and a coercive field at low temperatures up to $\approx 10$ kOe.[30,31] The SLs with thicker layers of LCO, i.e. L≥10 nm (2L>20 nm), show similar behavior to a 40 nm thin film of LCO on STO, with a slight decrease of $T_C$ and saturation magnetization (Figure 3 a,b). However, the magnetic behavior changes at L≤10 nm, with a strong reduction of the saturation magnetization and coercive field (Figure 3 c), and the appearance of a (probably antiferrro) magnetic signal around ≈68 K.

Zhang et al.[32] reported a change in the oxygen vacancy pattern for LCO on STO thin films thicker than 5 nm, signaling a change in the mechanisms of relaxation of epitaxial stress at this critical thickness. Also, Zhang et al.[33] found a suppression of the characteristic ferromagnetic phase of strained LCO in LCO/STO SLs with LCO layers thinner than ≈6 nm.

Thus, our results, particularly the magnetic signal at 68 K, point towards the existence of an additional interlayer region for L≤ 10 nm, whose composition cannot be determined from our data, but could be a mixed phase of the type $(Sr,La)(Co,Ti)O_x$. Several magnetic oxides of Sr-Co and Ti-Co are reported in the literature, like $Co_2TiO_4$ and $CoTiO_3$, with a smaller $T_C$ than LCO.[34]

Although, Ju *et al.*[35] showed that mixed interfaces could promote phonon transmission through a more gradual relaxation of epitaxial strain and acoustic mismatch, other types of defects, like oxygen vacancies or disordered cation substitution, could be however more detrimental for κ, as it seems to be the case here.

For an intuitive understanding of the effect of interfacial roughness, $\eta$, we defined a dimensionless parameter, *x*, which determines the fraction of coherent/incoherent phonons in the SL: $x = \frac{(1-\eta)}{L}$. Note that *x* decreases as L and/or $\eta$ increase, so that the effective thermal conductivity as a function of the periodicity, 2L, is obtained from the weighted fraction of coherent (ballistic) phonons, and incoherent phonons, following:

$$\frac{1}{\kappa} = x\frac{1}{\kappa_0} + (1-x)\left(\frac{1}{\kappa_0} + \frac{R_i}{L}\right) \qquad (2)$$

Although this is a very crude approximation (κ diverges at L→0), it gives an idea of the relevant parameters contributing to κ(L) of a SL. The κ(L) calculated for different values of the roughness is shown in Figure 4.

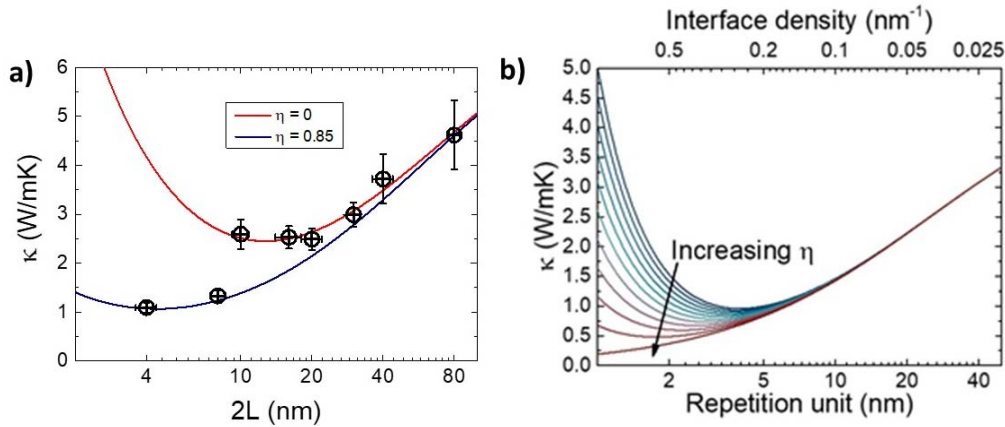

**Figure 4.** a) Fitting of experimental κ(150K) to equation (2), for different values of roughness b) Calculation of the thermal conductivity according to equation (2), as function of the interfacial roughness, and $\kappa_0$=4.5 Wm$^{-1}$K$^{-1}$.

Equation (2) fits the experimental κ(150K, L) data of the SLs down to 2L≈10 nm with a roughness $\eta \approx 0$ and $R_i = 3.57 \cdot 10^{-9}$ Km$^2$/W. The fitting also suggests that the plateau (or local minimum) of κ(L) around this region, could be consistent with a vanishing interfacial rugosity, which increases due to ionic interdiffusion as L decreases, according to magnetic data.

An increasing $\eta$ does not affect κ(L) at large L, where incoherent phonon transport is dominant, and the contribution of coherent phonons at low L becomes more relevant as $\eta$ decreases. The equation shows the existence of a minimum at the crossover between those regimes, if $\eta$ is small enough.

To probe the contribution of coherent phonons below/above the 2L≈10 nm boundary, we prepared additional sets of SLs with different total thickness but keeping their periodicity. In Figure 5b) we compare the thermal conductivity of a SL with 2L=40 nm (2L>$L_K$), and another one with 2L=8 nm (2L<$L_K$), varying the total thickness, $t$.

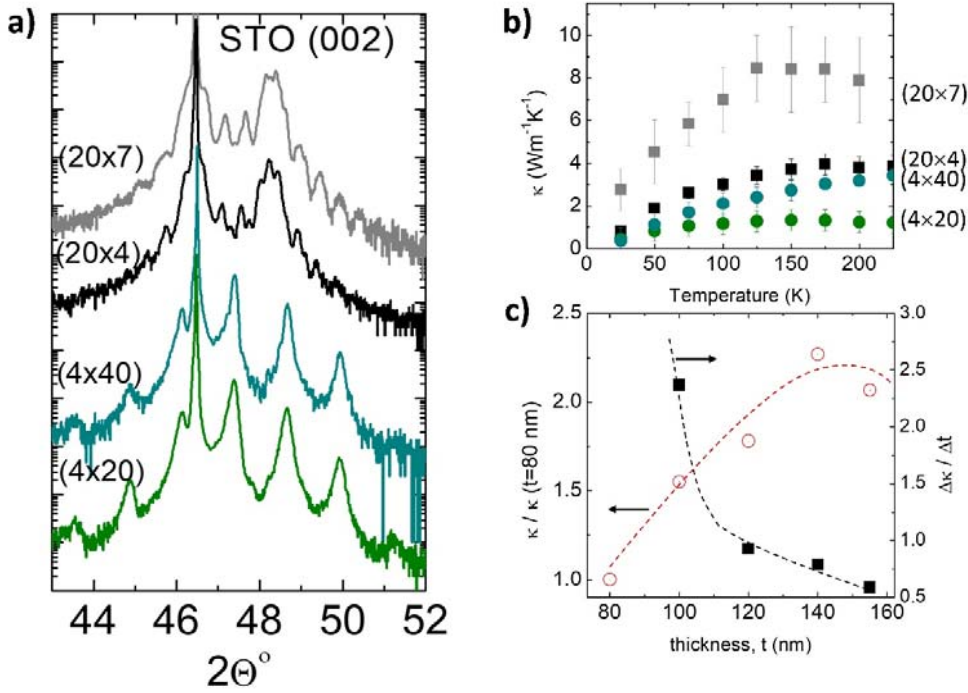

**Figure 5.** a) XRD of the SLs with L=4 nm and L=20 nm, with different total thicknesses. The period of the SL is maintained, as shown by the similar superlattice peaks in the XRD pattern. b) Temperature dependence of the thermal conductivity for the two sets of SLs. c) Thermal conductivity at 150K for the SLs, with different total thickness (open symbols), and relative increase of the thermal conductivity with respect to the 80 nm thick films, normalized by the relative increase of thickness: $\frac{\Delta k}{\Delta t} = \frac{[k-k(80\ nm)]/k(80\ nm)}{[t-80]/80}$. Lines are guides to the eye.

The results show that κ increases with $t$, at a similar rate, suggesting that, irrespective of the SL periodicity, there is a portion of phonons whose mean free path is limited by the total thickness of the SL and must be treated as coherent waves. On the other hand, the relative increase of κ, normalized by thickness of the SL, decreases continuously (Fig. 5c) and becomes less than ℓ at $t$≈120 nm. Beyond that point, increasing the SL thickness is not compensated by the contribution of larger mean free path phonons, and the probability of anharmonic phonon-phonon scattering increases sufficiently to reduce their contribution to the thermal conductivity. These results show that coherent phonons with a maximum ℓ≈120 nm contribute substantially to the thermal conductivity of LCO/STO SLs.

This opens the possibility to reduce the thermal conductivity of oxide SLs at large 2L, without introducing a large density of (or rough) interfaces, which is an interesting strategy for oxide-based thermoelectric devices. To probe this hypothesis, we prepared a SL with an average L=14 nm, but with an intentional aperiodicity of 15–20% between neighboring layers; (14×6):ap, see Figure 6. Long wavelength coherent phonons should be very much affected by a change in the periodicity of the SL,[35,36] while particle-like phonons should remain insensitive to it, as long as 2L>ℓ.

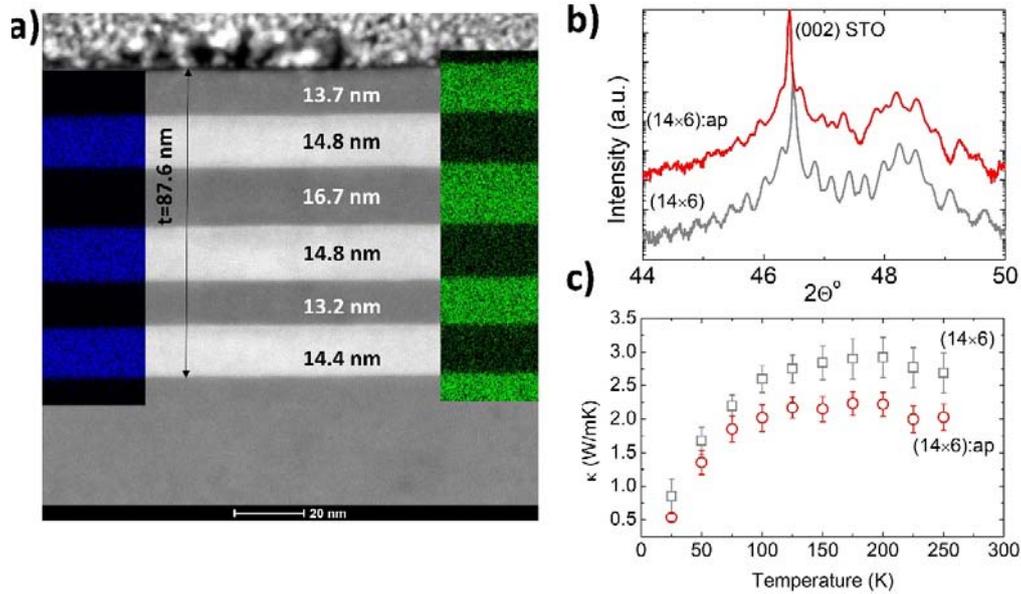

**Figure 6.** a) HAADF-STEM and EDX analysis of cross section lamellae of the aperiodic SL, (14×6):ap. The blue/green signal of the EDX map corresponds to Co/Ti, respectively. The X-ray diffraction pattern of the aperiodic and regular (14×6) SLs, is shown in b). The loss of periodicity is clearly reflected in the disappearance of many of the SL peaks on the (14×6):ap. c) Temperature dependence of the thermal conductivity of both SLs.

Despite the high quality of the interfaces, the X-ray diffraction pattern in Fig. 6b) shows the partial loss of the SL peaks in the aperiodic structure. The interferences between X-ray beams that produce SL peaks in the X-ray diffraction pattern have similar origin as the phonon wave interference, although a broad range of phonon frequencies contribute to κ. Similarly to that loss of periodicity, the thermal conductivity of the SL:ap is reduced by ≈25%, between 100-250 K. This result shows that even small variations in the periodicity may be a valid approach to control the thermal conductivity of SLs through the suppression of coherent phonons, even for a relatively small number of repetitions.

In summary, we have shown that the contribution of coherent phonons to oxide SLs is relevant in the whole period length and can be substantially reduced by small variations of the periodicity, without affecting its total thickness. Using this strategy, we tuned the thermal conductivity of $LaCoO_3/SrTiO_3$ SLs up to 20% at room temperature. This may have an interesting application in the development of low thermal conductivity devices, in which maintaining a relatively large thickness and clean interfaces is important for not deteriorating electrical transport, as in thermoelectrics.

## Materials and Methods

The $LaCoO_3/SrTiO_3$ (LCO/STO) SLs were deposited on (001)-oriented $SrTiO_3$ single crystals, used as received. The first layer of the SLs was always LCO, and the top layer was always STO. We have carefully optimized the deposition parameters to reduce as much as possible the presence of O and Sr vacancies in the STO films. The laser used for the PLD growth is a frequency-quadrupled Nd:YAG laser with a wavelength of 266 nm and maximum energy of 100 mJ. The fluence of the laser to grow the LCO and STO films was ≈ 2.3 $J/cm^2$. The substrate was kept at 725°C during deposition, and in an oxygen pressure of 200 mTorr. After deposition, the samples were cooled at 5°C/min in the same atmosphere. The growth rate (calibrated via ex-situ X-ray reflectivity analysis, for each material) with these conditions was 0.024 Å/pulse for STO and 0.027 Å/pulse for LCO. The thickness of each layer, and the total thickness of the SLs studied in this work, are reported in Table S1 of the supporting information.

HAADF-STEM and EDX analysis have been carried out in a probe- and image-corrected FEI Titan Themis microscope operated at 200 kV. Previously to HAADF-STEM and EDX analysis, a cross section lamella from each sample was prepared in a FEI Helios NanoLab 450S FIB-SEM.

The cross-plane thermal conductivity of the SLs was measured by the 3ω method; see supporting information for details of the measurements.

**Acknowledgments**

This work has received financial support from Ministerio de Economía y Competitividad (Spain) under projects No. MAT2016-80762-R, PGC2018–101334-BC21 and PID2019-104150RB-I00, Xunta de Galicia (Centro singular de investigación de Galicia accreditation 2019-2022, ED431G 2019/03), the European Union (European Regional Development Fund-ERDF) and the European Commission through the Horizon H2020 funding by H2020-MSCA-RISE-2016- Project No. 734187–SPICOLOST. E.L. is a Serra Húnter Fellow (Generalitat de Catalunya). D.B. acknowledges financial support from MINECO (Spain) through an FPI fellowship (BES-2017-079688). A.O.F. thanks MECD for the financial support received through the FPU grant FPU16/02572.

**Data and materials availability.** All data, and materials used in the analyses are available, under reasonable request.

## Supplementary Materials

Additional information about the preparation and structural characterization of the samples, as well as about the thermal conductivity measurements and analysis, is provided on the supplementary material accompanying this paper.

# Supporting Information for:

# Tuning Coherent-Phonon Heat Transport in LaCoO$_3$/SrTiO$_3$ Superlattices.

D. Bugallo,[1] E. Langenberg,[2] E. Carbó-Argibay,[3] Noa Varela Dominguez,[1] A. O. Fumega,[4,5] V. Pardo,[4] Irene Lucas,[6] Luis Morellón,[6] F. Rivadulla.[1,*]

[1]Centro de Investigación en Química Biolóxica e Materiais Moleculares (CIQUS) and Departamento de Química-Física, Universidade de Santiago de Compostela, 15782 Santiago de Compostela, Spain.

[2]Department of Condensed Matter Physics, Institute of Nanoscience and Nanotechnology (IN2UB), University of Barcelona, Spain.

[3]International Iberian Nanotechnology Laboratory (INL), Av. Mestre José Veiga s/n, 4715-330 Braga, Portugal.

[4]Departamento de Física Aplicada, Universidade de Santiago de Compostela, 15782 Santiago de Compostela, Spain.

[5]Department of Applied Physics, Aalto University, FI-00076 Aalto, Finland

[6]Instituto de Nanociencia y Materiales de Aragón (INMA), Universidad de Zaragoza and Consejo Superior de Investigaciones Científicas, 50009 Zaragoza, Spain.

## Pulser Laser Deposition of the superlattices.

The LaCoO$_3$/SrTiO$_3$ (LCO/STO) SLs were deposited on (001)-oriented SrTiO$_3$ single crystals, used as received. The first layer of the SLs was always LCO, and the top layer was always STO. The thickness of each layer, and the total thickness of the SLs studied in this work, are reported in Table S1.

**Table S1.** List of samples grown for this work. The nominal thickness of each layer, L, and total thickness, t, of the SLs synthesized for this work, as well as the experimental values obtained from different experimental methods.

| Sample | Layer thickness L (nm) | Total SL thickness t (nm) | Thickness XRD (nm) | Thickness XRR (nm) | Rugosity XRR (nm) |
|---|---|---|---|---|---|
| (2×40) | 2 | 80 | 1.9 | 1.96 | 0.5(1) |
| (4×20) | 4 | 80 | 3.8 | 3.8 | 0.8(2) |
| (4×40) | 4 | 160 | 3.8 | 3.9 | 0.5(1) |
| (5×16) | 5 | 80 | 5.4 | 5.3 | 0.7(3) |
| (5×24) | 5 | 120 | 5.7 | 5.6 | 0.9(3) |
| (8×10) | 8 | 80 | 9.0 | 7.8 | 1.1(3) |
| (10×8) | 10 | 80 | 10.5 | 10.4 | 0.7(2) |
| (14×6) | 14 | 84 | 17.5 | 15.2 | 1.4(2) |
| (14×6):ap | ≈13-16 | ≈ 88 (TEM) | | | |
| (20×4) | 20 | 80 | 20.9 | 21.5 | 1.3(2) |
| (20×6) | 20 | 120 | 19.6 | 19.5 | 1.5(3) |
| (40×2) | 40 | 80 | 40.2 | 46.1 | 0.9(1) |
| | | | | | |

## Ab-initio calculation of the cumulative thermal conductivity.

The ab-initio cumulative thermal conductivity is shown in Figure S1 for LCO and STO, at two different temperatures, vs the phonon mean free path, l$_{MFP}$. The STO two-step dependence of the cumulative lattice thermal conductivity is in agreement with previous calculations that highlight that thermal transport is carried not only by the acoustic modes, but also by the polar ones[1]. At 100 K, about 40% of heat is transported by ℓ<20 nm; this amount increases to ≈70 % at 240 K. Thus, between 100 K and room temperature, a periodic structure with a characteristic length 2L≈10-20 nm is expected to have a strong impact in the propagation of a large portion of phonons.

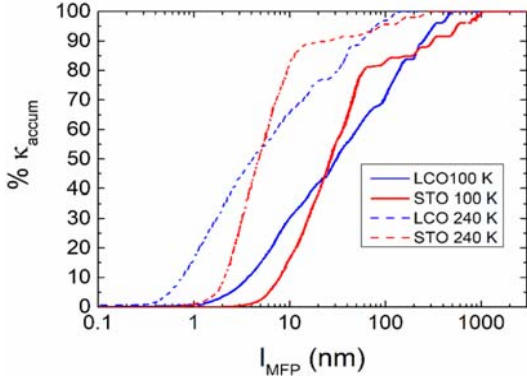

**Figure S1**. Cumulative thermal conductivity for SrTiO$_3$ and LaCoO$_3$, calculated ab initio.

The first principles density functional theory (DFT) calculations[2,3] were performed for LCO and STO in their low temperature structures. The generalized gradient approximation (GGA) in Perdew-Burke-Ernzerhof (PBE) scheme[4] was used for the exchange-correlation term with the PAW method as implemented in the VASP code[5]. The harmonic interatomic force constants (IFCs) were computed using the density-functional perturbation theory (DFPT)[6]. The third-order anharmonic IFCs were computed using the real-space supercell approach with a 3×3×3 supercell. These IFCs allow to compute the cumulative lattice thermal conductivity by iteratively solving the linearized Boltzmann-Peierls transport equation of phonons with the ShengBTE package[7]. Converged phonon momenta q-meshes of 15×15×15 were used for solving the transport equation of both compounds.

## Structural X-ray analysis of the SLs.

X-ray experiments were performed in a PANalytical Empyrean 4 circle diffractometer, equipped with a PIXCel$^{3D}$ detector and a Ge (220) double bounce monochromator, and a Cu source, 1.540598 Å (Cu K$_{\alpha 1}$). In the case of a SL, the X-ray beam reflects on the different interfaces of the multilayer, giving rise to constructive interferences at some angles that are related to the thickness of each layer:

$$2L = \frac{\lambda m}{2(sin(\theta_m) - sin(\theta))} \qquad \textbf{(S1)}$$

where $2L$ is the thickness of the repeating unit in the SL, $\lambda$ the wavelength, and $m$ is the order of the peak at $\theta_m$, with respect to the center one at $\theta$. The X-ray pattern of the SL formed by 8 repetitions of SrTiO$_3$/LaCoO$_3$, with L=5 nm of thickness in each layer (5×16), is shown in Figure S2, along with the fitting. The period obtained from these analyses is 2L=10.4 nm, close to the nominal one 2L=10 nm. The fit to equation (S1) is shown in Figure S2 also for a (20×4) SL.

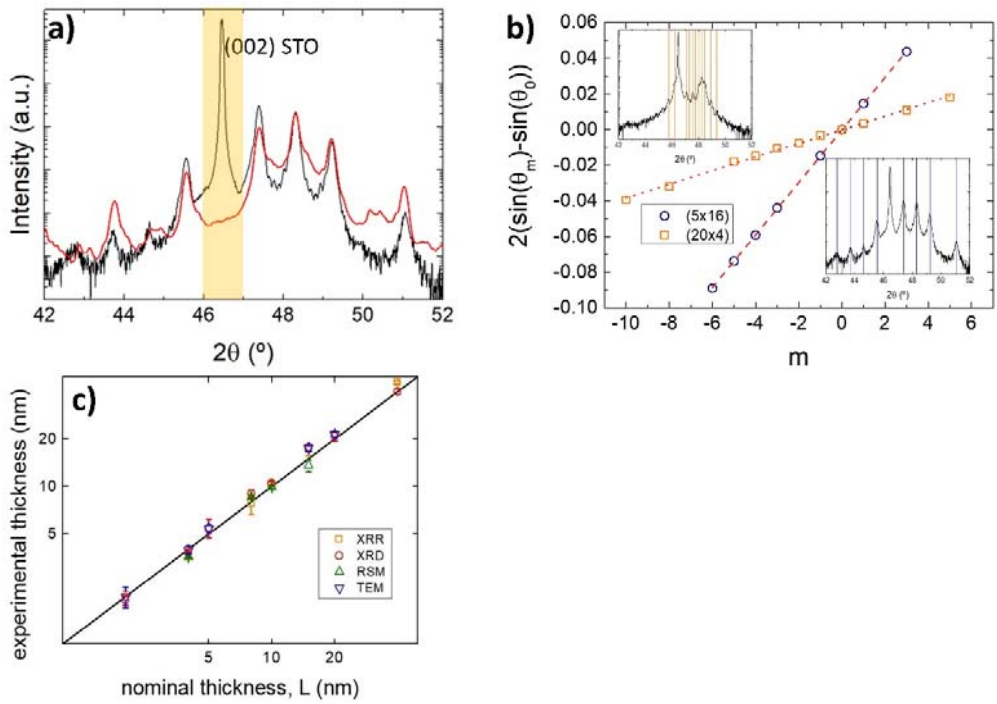

**Figure S2.** a) X-ray diffraction of a (5×16) SL, and simulation (red line); the peak of the substrate (shaded area) is not simulated. b) X-ray diffraction patterns with the position of the SL peaks marked (insets), and fitting to equation (S1), of the SL (20×4) and (5×16). c) Experimental vs. nominal thickness of each layer (L), obtained by different X-ray analyses, and TEM.

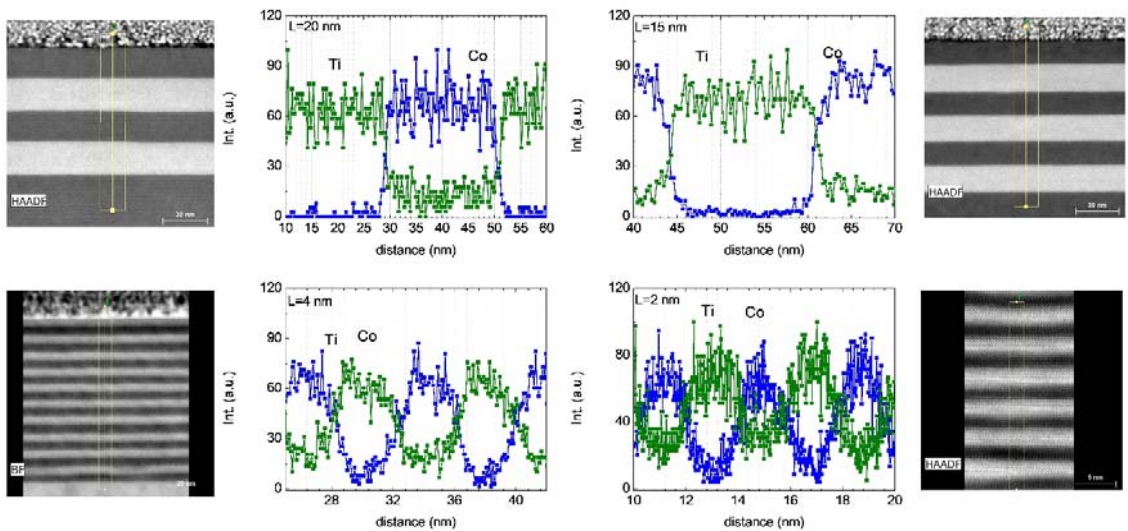

**Figure S3.** HAADF-STEM and EDX line-scan analysis of several SLs, with (from top, left) L=20 nm, 15 nm, 4 nm, and 2 nm.

# Experimental measurement of the thermal conductivity of the SLs by the 3ω method.

The cross-plane thermal conductivity of the SLs was measured by the 3ω method.[8,9] This is based on the use of a metallic resistor deposited on top of the sample to study, which acts, simultaneously, as a heater and a thermometer. An alternate current (of frequency ω) flows through the resistor, leading to a temperature oscillation in the film+substrate, due to Joule heating, which can be approximated, for a resistor of width w and with a penetration depth much smaller than the width of the resistor, by:

$$\Delta T \approx \frac{P}{\pi l \kappa} \left[ \frac{1}{2} \ln\left(\frac{4D}{w^2}\right) - \frac{1}{2}\ln(2\omega) + \ln(2) + \eta - \frac{i\pi}{4} \right] \quad \textbf{(S2)}$$

*P* is the power dissipated through the resistor, and η is the Euler-Mascheroni constant. With the change in temperature there will be associated a variation in the resistance of the metal heater at $2\omega$, as the resistance is not independent of temperature ($dR/dT \neq 0$). Applying Ohm's law taking into account that the resistance has a component that is a function of $2\omega$ and the current is a function of $\omega$, the voltage will have a component at $3\omega$, which will be directly related to the heating of the resistor:

$$V = I_0 \sin(\omega t)\, R = I_0 \sin(\omega t) \left[ R_0 + \frac{dR}{dT} \Delta T \cos(2\omega t + \phi) \right] = I_0 R_0 \sin(\omega t) + I_0 \frac{dR}{dT} \frac{\Delta T}{2} \sin(\omega t + \phi) - I_0 \frac{dR}{dT} \frac{\Delta T}{2} \sin(3\omega t + \phi)$$

$$\textbf{(S3)}$$

Thus, the third harmonic of the voltage ($V_{3\omega}$) is a direct probe of the heating of the sample, and therefore of its thermal conductance:

$$V_{3\omega} = \frac{I_0}{2} \frac{dR}{dT} \Delta T \quad \textbf{(S4)}$$

Taking this into account, from the slope ($m$) of $V_{3\omega}/V_{1\omega}$ vs $\ln(2\omega)$ plot, we can obtain the thermal conductivity of the sample:

$$\kappa = \frac{I_0^2}{4\pi l m} \frac{dR}{dT} \quad \textbf{(S5)}$$

A thin film between the resistor and the substrate can be modeled as a thermal resistance in series added to the system. Since the thin film's thickness, t, is much smaller than the thermal penetration depth and its thermal conductivity, $\kappa_{film}$, is smaller than that of the substrate, a frequency independent term adds to the heating:[10]

$$\Delta T_{film} = \frac{P \cdot t}{w \cdot l \cdot \kappa_{film}} \quad \textbf{(S6)}$$

In Figure S4 we show an example of an actual measurement of a substrate and film + substrate.

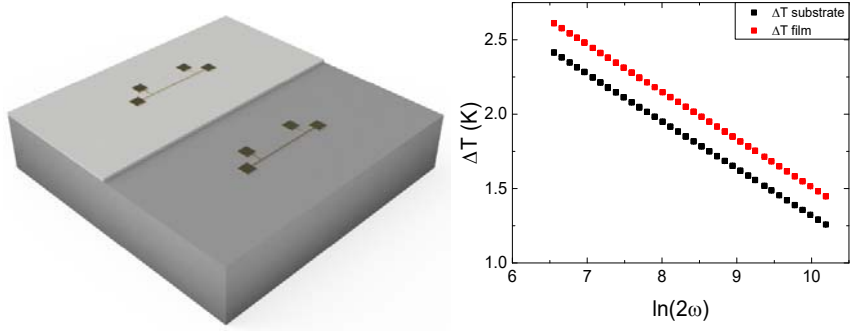

**Figure S4.** Left: Schematic of the resistances deposited on the film and substrate. Right: Measurement of the heating of a resistance deposited on top of an SrTiO$_3$ substrate (black) and another on top of a SL film, measured at 200 K. The thermal conductivity of the film is obtained from the offset between the two lines, according to equation S6.

For this work, the SLs were deposited by PLD covering only half the surface of the substrate; two resistors (10 nm Cr / 100 nm Pt; 1 mm long and 10 μm wide) were then deposited on the samples and on the film, so that the thermal conductivity of the two can be measured simultaneously.

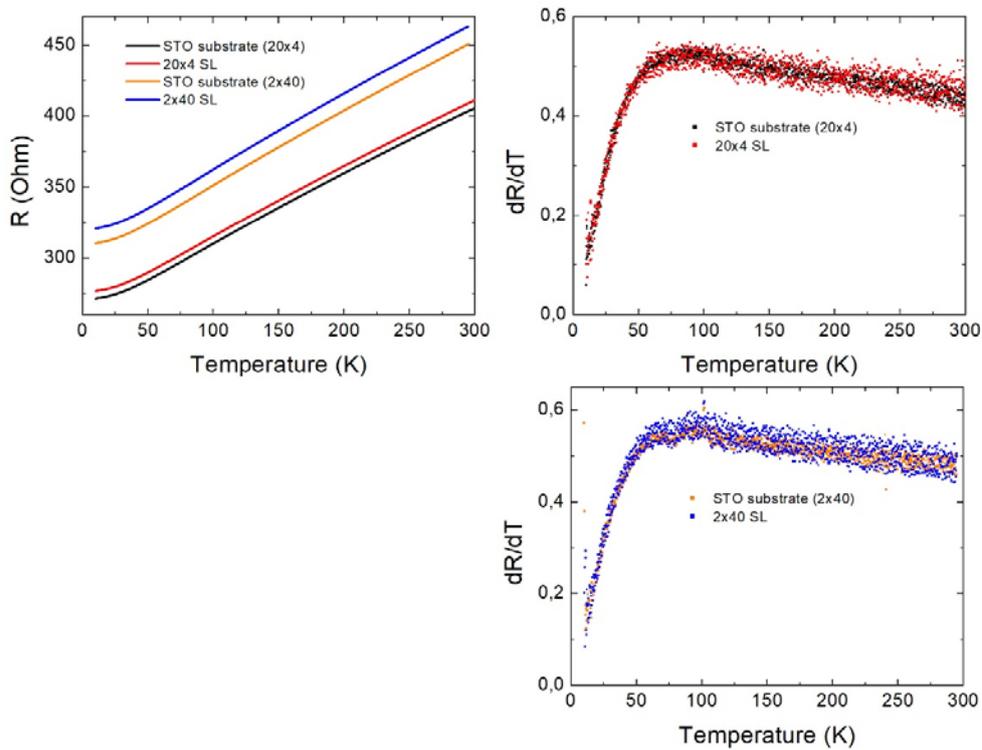

**Figure S5.** Temperature dependence of the electrical resistivity, and its derivative, for the resistances deposited on top of two SLs, and in the bare portion of the substrate.

LCO is not a perfect insulator; for that reason, as explained in the paper, our SLs always terminate in STO, which isolates the Pt sensor from the LCO.

Any leakage can be detected through a variation in the dR/dT of the Pt sensor. For that reason, we carefully checked this magnitude before every measurement of a SL. As you can see in figure S5, the R(T) and dR/dT of two Pt sensors on top of 2 SLs with different periodicity, as well as on top of a bare portion of the substrate, show the same behavior, within the error. This guarantees the absence of any measurable leakage through LCO, even with a 2 nm thick layer of STO (superlattice identified in the paper as SL 2x40).

To increase the accuracy, the first harmonic of the voltage was partially suppressed before entering the lock-in amplifier, using a specially designed circuit with three differential amplifiers and a set of resistors together with a variable resistor with PTC (positive temperature coefficient) close to zero.[11]

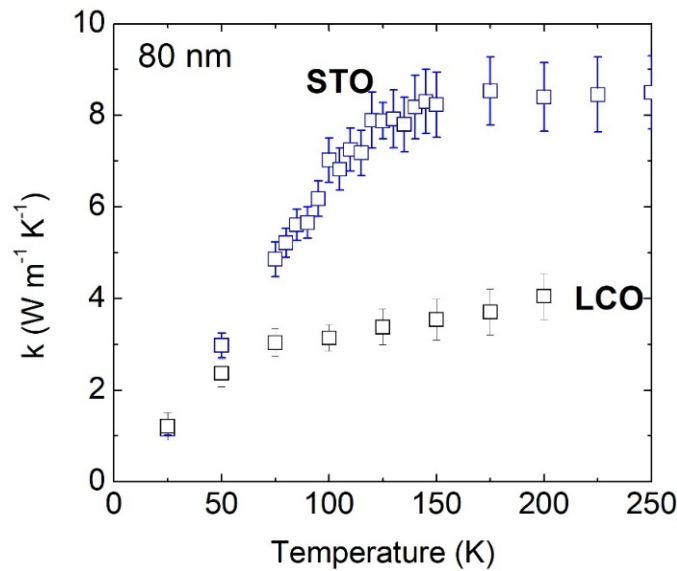

**Figure S6.** Temperature dependence of the thermal conductivity of 80 nm thick film of STO and LCO, respectively. To increase thermal contrast, the STO film is deposited on a MgO substrate.

We measured 80 nm thick films of pure LCO and pure STO (Figure S6). The 80 nm thick film of STO gives a value perfectly comparable with a film measured by Ravichandran et al.[12]; in this case synthesized by MBE and measured by time domain thermoreflectance. This supports the use of PLD and 3w for the synthesis and reliable measurement of the k(T) in oxide SLs.